\documentclass{iopart}

\usepackage{pslatex}
\usepackage{iopams}
\usepackage[dvips]{graphicx}

\begin{document}
\title[Accurate calibration of test mass displacement in the LIGO interferometers]{Accurate calibration of test mass displacement in the LIGO interferometers}
\author{E~Goetz$^1$, R~L~Savage~Jr$^2$, J~Garofoli$^2$\footnote{Current affiliation: Syracuse University, Syracuse, NY 13244, USA}, G~Gonzalez$^3$, E~Hirose$^4$\footnote{Current affiliation: Harvard-Smithsonian Center for Astrophysics, Cambridge, MA 02138, USA}, P~Kalmus$^5$\footnote{Current affiliation: California Institute of Technology, Pasadena, CA 91125, USA}, K~Kawabe$^2$, J~Kissel$^3$, M~Landry$^2$, B~O'Reilly$^6$, X~Siemens$^7$, A~Stuver$^6$ and M~Sung$^3$}
\address{$^1$ University of Michigan, Ann Arbor, MI 48109, USA}
\address{$^2$ LIGO Hanford Observatory, Richland, WA 99352, USA}
\address{$^3$ Louisiana State University, Baton Rouge, LA 70803, USA}
\address{$^4$ Syracuse University, Syracuse, NY 13244, USA}
\address{$^5$ Columbia University, New York, NY 10027, USA}
\address{$^6$ LIGO Livingston Observatory, Livingston, LA 70754, USA}
\address{$^7$ University of Wisconsin-Milwaukee, Milwaukee, WI 53201, USA}

\eads{\mailto{egoetz@umich.edu}, \mailto{savage\_r@ligo-wa.caltech.edu}}

\begin{abstract}
We describe three fundamentally different methods we have applied to calibrate the test mass displacement actuators to search for systematic errors in the calibration of the LIGO gravitational-wave detectors. The actuation frequencies tested range from 90 Hz to 1 kHz and the actuation amplitudes range from $10^{-6}$ m to $10^{-18}$ m. For each of the four test mass actuators measured, the weighted mean coefficient over all frequencies for each technique deviates from the average actuation coefficient for all three techniques by less than 4\%. This result indicates that systematic errors in the calibration of the responses of the LIGO detectors to differential length variations are within the stated uncertainties.
\end{abstract}


\section{Introduction}
The current generation of Earth-based interferometric gravitational wave (GW) detectors are sensitive to differential length variations with amplitude spectral densities as small as $10^{-19}$~m$/\sqrt{\textrm{Hz}}$.~\cite{LIGODetector,VirgoDetector,GEODetector,TamaDetector} The next generation of detectors are designed to be about a factor of ten more sensitive and in several cases they are either in the advanced planning stages~\cite{VirgoUpgrade,LCGT} or have already begun construction~\cite{AdLIGOref,GEOUpgrade}. In order to take full advantage of the scientific reach afforded by these detectors, continuous calibration with accuracy and precision approaching the 1\% level will be required.~\cite{ReqRespErrors} Thus, detector calibration is an active area of research for all GW projects, with a significant percentage of resources being devoted to calibration-related activities. These activities include amplitude and phase calibration in both the frequency domain~\cite{OcalPaper,VirgoCal,GEOCal,TamaCal1} and in the time domain~\cite{LIGOhoft,PcalTiming,Virgohoft,GEOhoft,TamaCal2}.

Historically, the LIGO project has relied on a calibration method that requires extrapolation from test mass displacements that are about twelve orders of magnitude larger than the expected apparent displacements that will be caused by gravitational waves.~\cite{S1paper} While the precision of this method has improved over the more than five years it has been employed, the possibility of a large systematic error has not been eliminated until now.

To search for systematic errors in the LIGO calibration procedure, we have employed three fundamentally different actuator calibration techniques: the {\it free-swinging Michelson} method~\cite{S1paper} that relies on the wavelength of the laser light in the interferometer, the {\it photon calibrator} method~\cite{LIGOPcal} that uses the recoil of photons from an auxiliary laser source to induce calibrated test mass displacements with amplitudes close to the detector sensitivity limit, and the {\it frequency modulation} method~\cite{VCOpaper} that relies on the frequency-to-length transfer function for a Fabry-Perot optical cavity. Our investigations have spanned the frequency range from 90 Hz to 1 kHz and the range of actuation amplitudes from $10^{-6}$ m to $10^{-18}$ m.

The closed-loop calibration of the LIGO displacement actuators is described in the following section, and an overview of each of the three calibration methods is presented. Measurement results and estimates of the uncertainties associated with each method are given in section~\ref{sec:MeasResultsUncert}. Conclusions and implications for future calibration work are discussed in section~\ref{sec:conclusions}.

\section{Calibration methods}\label{sec:calMethods}
\begin{figure}
	\begin{flushright}
		\includegraphics[width=0.80\textwidth]{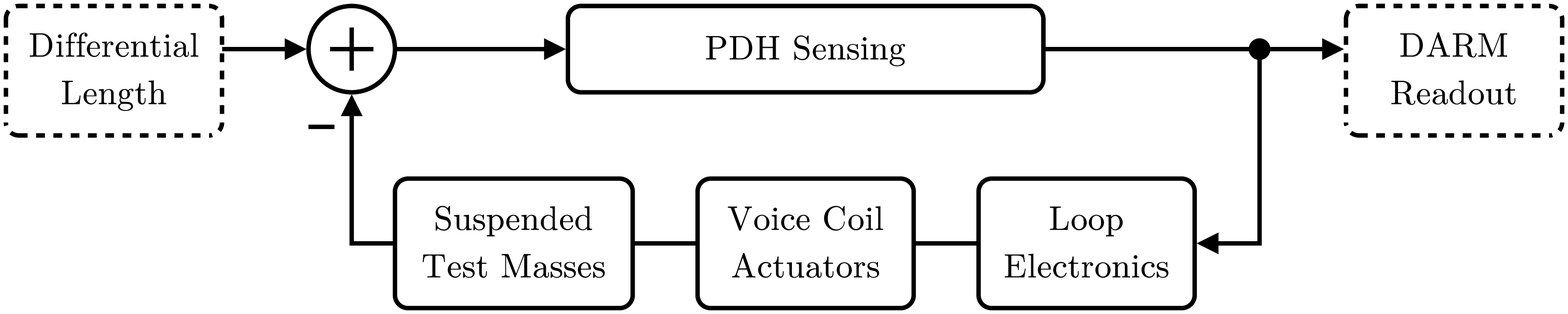}
	\end{flushright}
	\caption{Schematic diagram of the differential arm length (DARM) feedback control loop.  The Pound-Drever-Hall sensing technique is used to produce the DARM readout signal.  This signal is amplified and filtered, then directed to the voice coil actuators which displace the end test masses to maintain the resonance condition in the interferometer.}
	\label{fig:darm}
\end{figure}

\begin{figure}
	\begin{flushright}
		\includegraphics[width=0.60\textwidth]{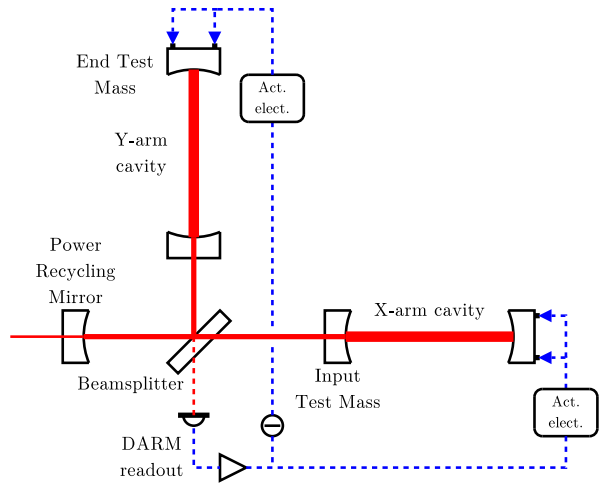}
	\end{flushright}
	\caption{Schematic diagram of the LIGO optical configuration showing the DARM feedback control loop that actuates on the ETMs. The difference in arm lengths is sensed at the anti-symmetric port by a photodetector using the Pound-Drever-Hall reflection locking technique. The output voltage is filtered and amplified then directed to the voice coil actuators for each end test mass.}
	\label{fig:IFO}
\end{figure}

The LIGO detectors are power-recycled Michelson interferometers with Fabry-Perot arm cavities.~\cite{LIGODetector} The interferometer optics are suspended as pendulums and actively controlled to maintain resonance conditions. Changes in the difference between the lengths of the interferometer's arms, the degree of freedom most sensitive to gravitational waves, are sensed at the anti-symmetric port. The differential arm length (DARM) control system uses a variation of the Pound-Drever-Hall (PDH) radio-frequency locking technique~\cite{PDHpaper,LIGOPDH} with electromagnetic displacement actuators consisting of voice coils interacting with magnets glued to the back surfaces of the mirrors. The idealized force-to-length transfer function is proportional to the inverse of the square of the excitation frequency. The DARM control loop is shown schematically in figure~\ref{fig:darm}, and in the context of the interferometer in figure~\ref{fig:IFO}. The DARM readout signal is amplified, filtered, and then directed to the voice coil actuators on the mirrors at the ends of the arm cavities. These mirrors, together with the input arm cavity mirrors, are the {\it test masses} for gravitational wave signals.

Reconstruction of differential length disturbances from the DARM readout signal requires correcting for the DARM closed-loop response. Measurement of the overall DARM loop transfer function is relatively straightforward. However, measurement of the transfer functions for components of the loop such as the sensing and actuation paths is more difficult. Calibrating the response of the interferometer to differential length variations is directly dependent on the characterization of the actuation function. In addition to the voice coil actuators and the suspended test masses, the actuation path includes actuation electronics that convert the drive voltages to coil currents. These electronics have two modes of operation, a low-noise {\it Run} mode, and an {\it Acquire} mode that allows larger drive amplitudes. Calibration of this actuation chain is the focus of this article.

\subsection{Free-swinging Michelson}\label{sec:fsm}
The technique that has been the traditional calibration method employed by LIGO is referred to as the free-swinging Michelson method because it relies on measurement of Michelson interference fringes when the suspended optics are swinging freely. It thus uses the wavelength of the interferometer's laser light as a length reference, and to calibrate the end test mass (ETM) actuation function via a series of measurements made with both the interferometer and the actuation path electronics in various configurations. The required drive amplitudes are on the order of $10^{-12}$ m.  This process is described in detail in \cite{S1paper}; an overview of the procedure is given here.

\begin{figure}
	\begin{flushright}
		\includegraphics[width=1.0\textwidth]{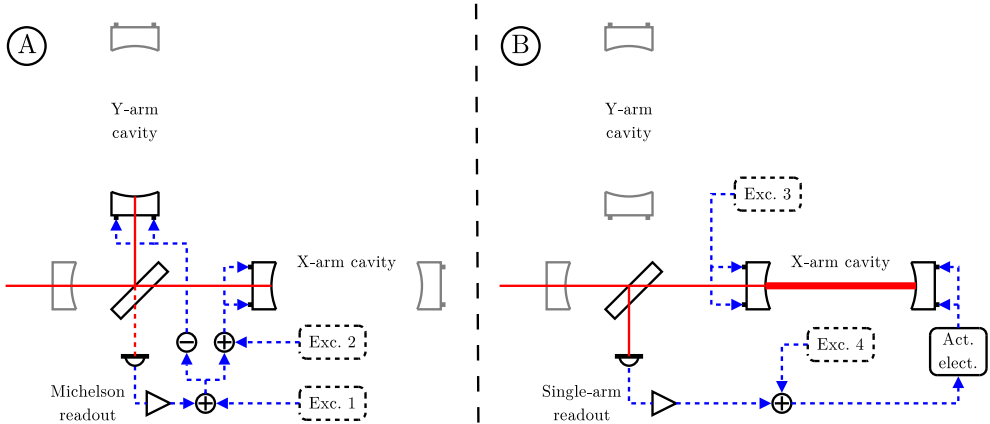}
	\end{flushright}
	\caption{A: Schematic diagram of the simple Michelson configuration with misaligned optics shown in grey. The electronics are configured to feed back to the ITMs to maintain the dark-fringe condition at the anti-symmetric port. B: Schematic diagram of the single arm lock configuration, again with misaligned optics shown in grey. The electronics are configured to feed back to the ETM to maintain the resonance condition.}
	\label{fig:FSM}
\end{figure}

The first step in this process is to misalign the ETMs and the power recycling mirror, then align the input test masses (ITMs) and beam splitter to form a simple Michelson interferometer as shown in figure~\ref{fig:FSM}A.  The length control servo electronics are configured to lock this setup on a dark fringe, i.e. with destructive interference of the light from each arm at the anti-symmetric port where the photodetector is located. Two transfer function measurements are made in this configuration, the overall open-loop transfer function (Exc. 1 in figure~\ref{fig:FSM}) and the transfer function from actuation of an ITM (Exc. 2) to the Michelson readout signal. The second measurement is repeated for the other ITM. Then, the feedback control is switched off, allowing the optics to swing freely in response to the seismic motion filtered by the vibration isolation systems. With the loop unlocked, the time series of the photodetector output is recorded as the Michelson length difference changes, causing the output to vary between the bright-fringe and dark-fringe levels.  The difference between the maximum and minimum outputs corresponds to relative ITM motion of one-fourth of the wavelength of the laser light, thus providing a calibration of the anti-symmetric photodetector output signal in this Michelson configuration.  Combining this result with the transfer functions resulting from ITM actuation and the overall open-loop transfer function yields a calibrated actuation function for the ITMs.

The next step in the process is to misalign one of the ITMs and realign the ETM on the opposite arm. The servo electronics are then configured to feed back to the position of the ETM forming a resonant Fabry-Perot arm cavity as shown in figure~\ref{fig:FSM}B. Transfer functions from ITM and ETM actuations (Exc. 3 and 4) to the single-arm readout are then measured. Their ratio, combined with the calibration of the ITM actuation function yields the ETM actuation function. Making similar measurement with only the other interferometer arm cavity aligned yields the calibrated actuation function for the other ETM.

The swept-sine measurements that are performed in the single-arm configuration require the actuation path electronics to be in the Acquire mode. However, in the Acquire mode the coupling of electronics noise to test mass displacement is 3-4 orders of magnitude larger than in the Run mode, which is used for GW searches. There are four parallel paths for the four voice coil actuators on each ETM and several components in each path. Measuring and combining the electronics transfer functions with overall precision approaching 1\% is a task that has proved difficult to achieve. Converting calibrations performed using the free-swinging Michelson method to the high-sensitivity, fully-locked interferometer configuration requires correcting for differences between the Acquire and Run mode actuation paths.

\subsection{Photon calibrator}\label{sec:pcal}
The photon calibrator method uses an auxiliary, power-modulated laser to induce calibrated displacements via the recoil of photons from the surface of the ETM. Photon calibrators have been used before~\cite{GlasgowPcal,GEOPcal} and are currently being employed at several GW observatories~\cite{LIGOPcal,VirgoPcal,GEOcompare}. A detailed description of the LIGO photon calibrators can be found in \cite{LIGOPcal}.

\begin{figure}
	\begin{flushright}
		\includegraphics[width=0.80\textwidth]{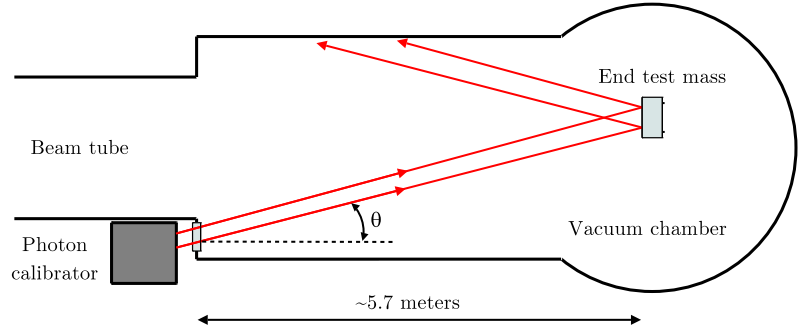}
	\end{flushright}
	\caption{Schematic diagram of a LIGO photon calibrator with output beams reflecting from an end test mass inside the vacuum envelope.}
	\label{fig:PcalAndVacuumChamber}
\end{figure}

To avoid sensing the region of the ETM that is elastically deformed by the photon pressure~\cite{HildEffect}, the LIGO photon calibrators have evolved to a two-beam configuration with the beams symmetrically displaced about the center of the ETM as shown in figure~\ref{fig:PcalAndVacuumChamber}. The spot locations and powers in the two beams are balanced to minimize induced rotations which, for a mis-centered interferometer beam, would cause apparent length variations. The total sensed test mass motion is given by
\begin{equation}
\Delta L (f)  = -\frac{P\cos\theta}{2\pi^2Mcf^2} \left[1+ \frac{(\vec{a}\cdot\vec{b})M}{I}\right]
\end{equation}
where $f$ is the excitation frequency, $P$ is the modulated laser power reflecting from the ETM, $\theta$ is the angle of incidence, $M$ is the mass of the ETM, $c$ is the speed of light, $\vec{a}$ and $\vec{b}$ are the weighted-average photon calibrator and interferometer beam centering offsets and $I$ is the moment of inertia of the ETM. In LIGO, the typical power-modulation amplitude is about 100 mW, the laser wavelength is 1047 nm, the angle of incidence is about 10 degrees, and the mass of the ETM is about 10.3 kg. Typical beam offsets are less than a few cm, so the rotation-induced apparent displacement is more than 100 times smaller than the longitudinal displacement. To minimize bulk elastic deformation caused by the photon calibrator beams, they are displaced from the ETM center by about 0.65 times the radius of the ETM, where finite-element modeling indicates that the coupling to bulk deformation sensed by a centered interferometer beam is minimized.~\cite{TMdeform}

Accuracy of calibration using the photon calibrators is directly dependent on estimates of the absolute power reflecting from the ETM. Temperature-stabilized InGaAs photodetectors mounted on integrating spheres and calibrated at NIST (National Institute of Standards and Technology) are employed to calibrate an internal power sensor. They are also used to measure the transmissivities of the vacuum windows and the reflectivities of the ETMs. The internal sensor provides a continuous monitor of the photon calibrator output power. Calibration of the ETM actuation coefficient at a given frequency is achieved by modulating the ETM position using the voice coil actuators at the desired frequency and simultaneously modulating the laser power at a nearby frequency, typically offset by about 0.1 Hz. The ratios of the amplitudes of the lines induced in the amplitude spectral density of the DARM readout signal combined with the calculated ETM motion caused by the photon calibrator yields a calibration of the voice coil actuation path. This calibration is to first order insensitive to changes in the interferometer operating conditions because the frequencies of the excitations are closely spaced and the measurements are made simultaneously. These measurements are performed with the interferometer locked in the configuration that is used during GW searches and with actuation forces inducing displacements that are close to the sensitivity limit.

\subsection{Frequency modulation}\label{sec:freqMod}
Frequency modulation provides a force-free method for calibrating the ETM voice coil actuators. By modulating the frequency of the laser light in a single-arm configuration, we create an effective length modulation detected by the PDH sensor. The magnitude of the induced effective length modulation is given by the dynamic resonance condition for a Fabry-Perot cavity~\cite{DynamResonance},
\begin{equation}
C(f) \frac{\Delta \nu(f)}{\nu} = -\frac{\Delta L(f)}{L}
\end{equation}
where $C(f)$ is the normalized frequency-to-length transfer function, $\nu$ is the laser frequency, $L$ is the length of the arm cavity, and $f$ is the frequency of the modulations, $\Delta\nu$ and $\Delta L$.  For the frequencies considered in our measurements, much lower than the cavity free spectral range, $C(f) \simeq 1$ such that 
\begin{equation}
\Delta L(f) \simeq - L \frac{\Delta \nu(f)}{\nu}.
\end{equation}

\begin{figure}
	\begin{flushright}
		\includegraphics[width=1.0\textwidth]{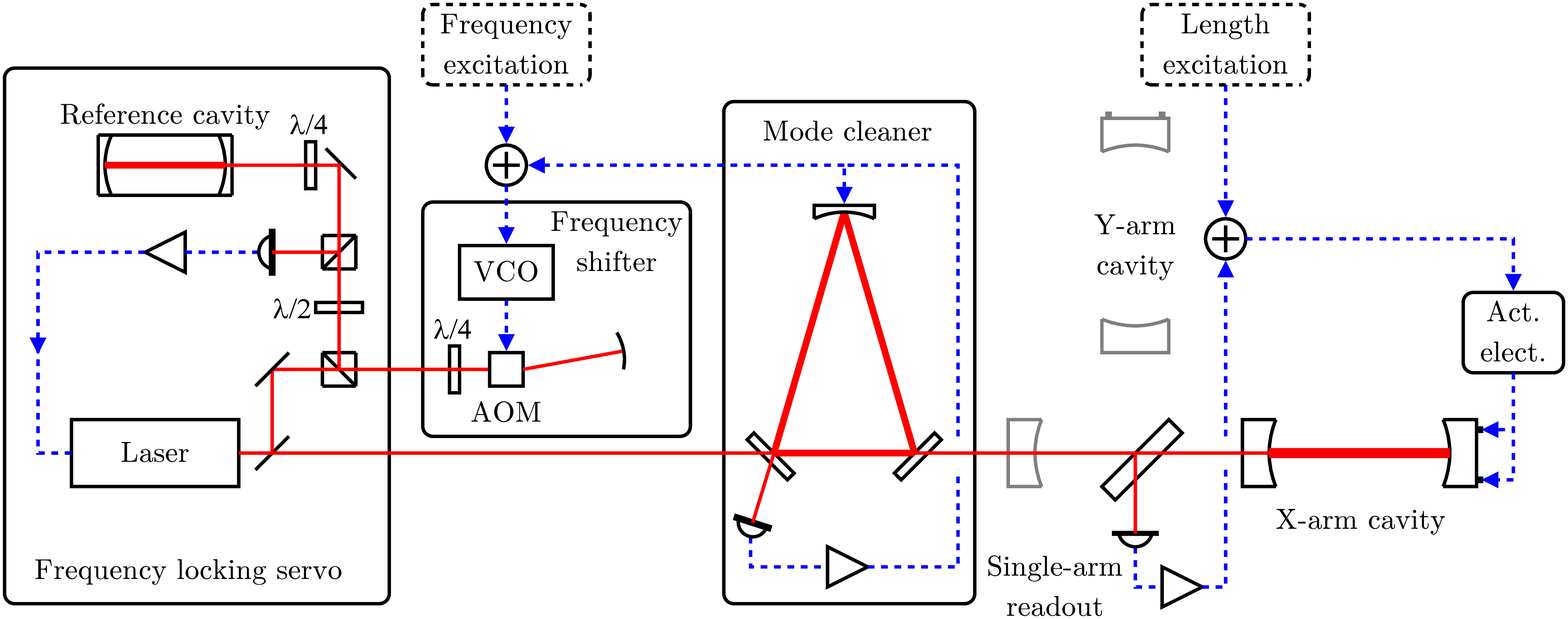}
	\end{flushright}
	\caption{Schematic diagram of the frequency modulation method.  The laser frequency is locked to the resonance of the reference cavity. The VCO injects a frequency modulation into the frequency locking servo loop via the double-passed acousto-optic modulator (AOM). The frequency servo acts on the laser frequency to cancel the injected modulation, thus imposing the inverse of the modulation on the laser light directed to the mode cleaner. The mode cleaner acts as a low-pass filter for the frequency modulation. Measurements are made with the interferometer in the single-arm configuration (misaligned optics shown in grey).}
	\label{fig:VCO}
\end{figure}

As described in more detail in \cite{VCOpaper} and shown schematically in figure~\ref{fig:VCO}, the frequency modulation is applied via a frequency shifter that is embedded in the frequency stabilization servo that is used to lock the laser frequency to a reference cavity. The frequency shifter uses a voltage-controlled oscillator (VCO) that drives a double-passed acousto-optic modulator to modulate the frequency of the laser light directed to a three-mirror suspended mode cleaner. Calibrating the frequency shift induced by a sinusoidally-varying control voltage is a key step in applying this technique. In practice, we calibrate the VCO by locking it to a frequency synthesizer that is itself locked to a frequency standard. The frequency modulation induced by an applied offset is determined by measuring the frequency modulation sideband-to-carrier ratio using a spectrum analyzer.

The frequency-modulated light passes through the mode cleaner, which acts as a low-pass filter for frequency fluctuations. The mode cleaner filtering is measured via a power-modulation transfer function using photodetectors located both upstream and downstream of the mode cleaner cavity. This filtering is well approximated by a single pole at a frequency of about 4 kHz.

As with the photon calibrator technique, we generate two peaks in the spectrum of the single-arm readout signal, in this case by simultaneously modulating the position of the ETM using the voice coil actuators while applying the frequency modulation at a frequency separated by about 0.1 Hz. Excitation amplitudes are on the order of $10^{-12}$ to $10^{-14}$ m, with the lower bound for frequency modulation dictated by the mode cleaner control signal background at the input to the VCO and the upper bound for length modulation governed by saturation concerns in the Run mode actuation path electronics. Comparing the heights of the peaks induced in the single-arm readout signal and multiplying by the VCO calibration yields the calibration of the voice coil actuation. Unlike the free-swinging Michelson transfer functions which are typically measured at many frequencies with large drive amplitudes, the frequency modulation measurements use long integration times at a single frequency with the voice coil actuation electronics in the Run mode. The difficult step of precise compensation for the differences in Run versus Acquire electronics paths is therefore not required.

\section{Measurement results and uncertainties}\label{sec:MeasResultsUncert}
The measurements reported here were performed at the LIGO Hanford Observatory (LHO) during a period dedicated to calibration-related activities at the end of the S5 science run~\cite{LIGODetector} in October and November of 2007.  Two interferometers were operating at LHO, the H1 interferometer with 4-km-long arm cavities and the H2 interferometer with 2-km-long arms.  The photon calibrators for three of the four ETMs had already been converted to two-beam configurations.  Only the H1 y-arm system was still in a one-beam configuration with the beam centered on the ETM.

The photon calibrator and frequency modulation measurements were similar in that they were made at a few discrete frequencies.  Long integration times ($\sim$250 seconds) and multiple averages enabled assessment of the statistics for the measurements and reduction of the standard errors.  On the other hand, swept-sine measurements with approximately fifty measurement frequencies between 90 Hz and 1 kHz were used for the free-swinging Michelson technique.  This is not a fundamental requirement for this method, rather it is the type of measurement found to be most useful for estimating the actuation function over the relevant band of frequencies.  We thus compare swept-sine measurements for the free-swinging Michelson method with single-frequency measurements at a few discrete frequencies for the other two methods.

\begin{figure}
	\begin{flushright}
		\includegraphics[width=1.0\textwidth]{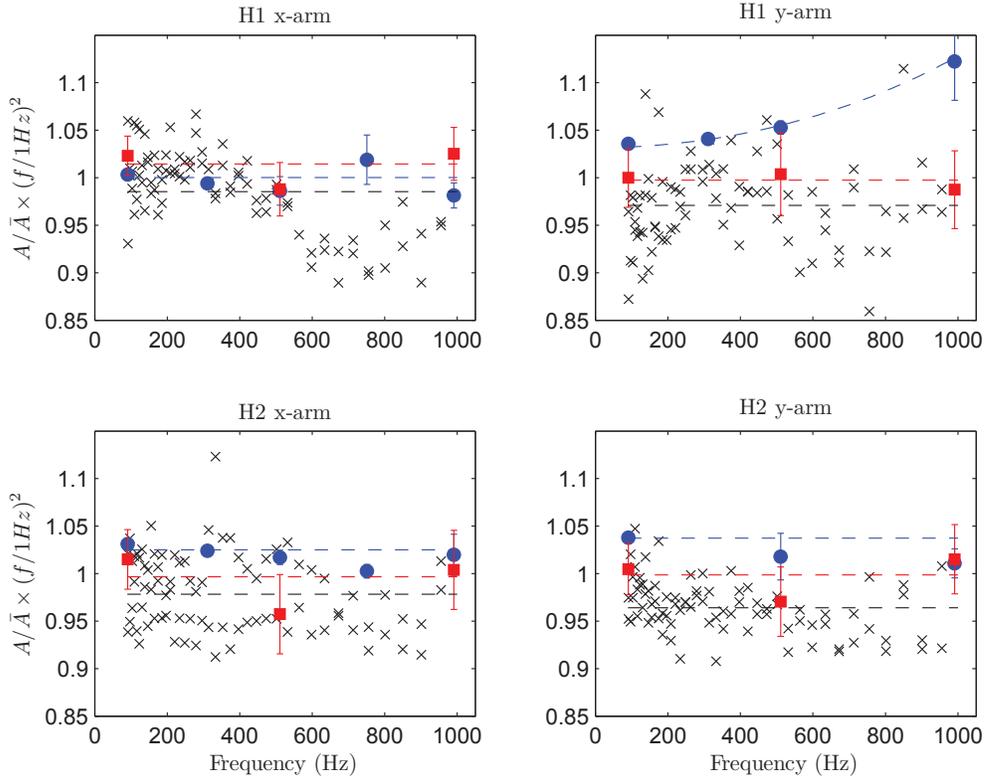}
	\end{flushright}
	\caption{Comparison of ETM actuation coefficients measured with three techniques: free-swinging Michelson (black crosses), photon calibrator (blue circles), and frequency modulation (red squares). The data are multiplied by the square of the measurement frequency and normalized to the average of the three weighted mean values (dashed horizontal lines) for each method, $\overline{A}$. The free-swinging Michelson data are plotted without error bars; for visibility, $3\sigma$ statistical error bars are plotted for the other two methods. The single-beam H1 y-arm photon calibrator data show the influence of local elastic deformation by photon radiation pressure.}
	\label{fig:compare}
\end{figure}

The results of measurements carried out with all three techniques are plotted together in figure~\ref{fig:compare}.  For each datum, the calculated actuation coefficient is multiplied by the square of the measurement frequency to facilitate comparison with a simple $f^{-2}$ functional dependence. The free-swinging Michelson data, plotted without error estimates, are from two separate calibration sequences that were carried out on consecutive days. For better visibility, the error bars for the photon calibrator and frequency modulation data are $\pm 3 \sigma$ estimates of statistical uncertainties only, with $\sigma$ a representative standard error for the averaged measurements. Estimates of systematic uncertainties for each calibration method~\cite{S5paper,LIGOPcal,VCOpaper} have intentionally been omitted so that the overall systematic uncertainty can be evaluated by comparing the results of the three methods with statistical precision indicated by the error bars, or the scatter in the data for the free-swinging Michelson results. For each ETM, the dashed horizontal lines indicate the simple mean value for the free-swinging Michelson method and the weighted mean values for the photon calibrator and frequency modulation methods. All data are normalized to the average of the mean values for the three methods, $\overline{A}$.  The H1 y-arm photon calibrator is in a single-arm configuration, therefore local elastic deformation caused by the photon calibrator forces and sensed by the interferometer beam causes the overall actuation function to rise dramatically with increasing frequency.~\cite{LIGOPcal}  We thus fit a curve with the functional form predicted by Hild, et al.~\cite{HildEffect} to the data in figure~\ref{fig:compare} instead of averaging over frequencies. The low frequency asymptote is the actuation coefficient we would expect for a two-beam photon calibrator operating on this mass.

The data for all three measurement methods exhibit variations with frequency that appear to be inconsistent with a simple $f^{-2}$ frequency dependence. This is likely due to either frequency-dependent systematic errors in the calibration methods or frequency-dependent variations in the actuation path electronics in Run mode. For the free-swinging Michelson data, the sensing function and errors in making the Run/Acquire correction can also introduce frequency-dependent variations.

Potential sources of systematic errors for the free-swinging Michelson technique include time-dependent alignment variations during sequential measurements, measuring in different interferometer and electronics configurations, and extrapolation over nearly 12 orders of magnitude in actuation range. Overall systematic uncertainty, estimated from the observed spread in the data and from propagating errors through the many steps of the technique, is approximately 10\%.~\cite{S5paper} The primary identified sources of systematic uncertainty for the photon calibrator method are rotation due to beam centering offsets and absolute power calibration. With reasonable centering tolerances and the demonstrated power calibration accuracy, the estimated total systematic uncertainty can be reduced to the order of 1\%.~\cite{LIGOPcal} For the frequency modulation method, the dominant sources of potential systematic errors are the calibration of the VCO which relies on accurate measurement of the sideband-to-carrier ratios and extrapolation over approximately 5 orders of magnitude in actuation range. In practice, the overall estimated uncertainty for this method due to known sources of systematic errors can also be reduced to the 1\% level.~\cite{VCOpaper} 

All calculated actuation coefficients--for all frequencies, for all masses, and for all three techniques--fall within the $\pm15$\% ranges plotted in figure~\ref{fig:compare}. The maximum difference between the mean values over all frequencies of any two methods for any ETM is less than 8\%. The maximum difference between the mean value for any method and the average of the mean values for all three methods, $\overline{A}$, for any ETM, is 3.7\%; the standard deviation of the twelve differences (all four ETMs) is 2.4\%.

\section{Conclusions}\label{sec:conclusions}
Errors in the measurement of the actuation functions directly translate into errors in the inferred responses of the LIGO detectors to length variations. Potential errors in the determination of the actuation functions have been the principal concern regarding overall detector amplitude calibration uncertainty. We have presented the results of measurements made to compare three intrinsically different test mass actuator calibration methods in order to bound potential systematic errors in the free-swinging Michelson method, the main calibration technique traditionally used by LIGO. The observed level of consistency, in light of the differences in the techniques--from fitting interference fringes in a Michelson configuration to photon pressure actuation to laser frequency modulation, in the methods of application--from Michelson and single-arm configurations to the fully-locked configuration used during GW searches, and in the range of actuation--from $10^{-6}$ m to $10^{-18}$ m, indicates that the actual voice coil actuation functions are within the bounds of these measurement results. This, together with the independent measurement of detector displacement sensitivity afforded by the photon calibrator, gives us confidence that the LIGO detector calibration is within the stated uncertainty estimates.~\cite{S5paper}

As we have described above, calibration of the displacement actuators of an interferometric GW detector can be a tedious and complicated enterprise. Future GW detectors will have even more sophisticated actuation and readout chains. With optimal signal extraction requiring overall calibration accuracies of 1\% in amplitude, plus constraints on phase and timing, continued improvement of calibration methods and procedures will be imperative. For instance, ongoing finite-element modeling of actuation forces interacting with LIGO-style test masses indicates that calibration at the 1\% level will require correcting for the bulk deformation induced by the actuation forces. This is particularly relevant at frequencies near and above 1 kHz. Techniques such as the photon calibrator method that can provide on-line calibration of the DARM readout signal during GW searches will be particularly attractive. Multiple parallel calibration efforts will likely be required to achieve these goals. Our experience has exposed the benefits of employing several different methods to search for and mitigate potential sources of systematic errors. The need for precise and accurate on-line calibration dictates that calibration requirements must be an integral part of the design of gravitational-wave detectors to achieve their full scientific potential.

\ack We gratefully acknowledge the support of the LIGO Scientific Collaboration and the LIGO Laboratory for construction and operation of the LIGO detectors. We also thank the National Science Foundation for support under grant PHY-0555406. LIGO was constructed by the California Institute of Technology and the Massachusetts Institute of Technology with funding from the National Science Foundation and operates under cooperative agreement PHY-0107417. This document has been assigned LIGO Laboratory document number P0900155.


\section*{References}
\begin{thebibliography}{99}
	\bibitem{LIGODetector} Abbott~B, \etal 2009 {\it \RPP} {\bf 72} 076901
	\bibitem{VirgoDetector} Acernese~F, \etal 2006 {\it \CQG} {\bf 23} S635--42
	\bibitem{GEODetector} L\"uck~H, \etal 2006 {\it \CQG} {\bf 23} S71--8
	\bibitem{TamaDetector} Takahashi~R and the TAMA Collaboration 2004 {\it \CQG} {\bf 21} S403--8
	\bibitem{VirgoUpgrade} The Virgo collaboration 2008 {\it Virgo doc.} VIR-0089A-08 (https://tds.ego-gw.it)
	\bibitem{LCGT} Kuroda~K and the LCGT Collaboration 2006 {\it \CQG} {\bf 23} S215--21
	\bibitem{AdLIGOref} Harry~G for the LIGO Scientific Collaboration 2009 to be submitted to {\it Amaldi 8 Conf. Proc.}
	\bibitem{GEOUpgrade} L\"uck~H, \etal 2009 submitted to {\it J. Phys.: Conf. Ser.}
	\bibitem{ReqRespErrors} Lindblom~L 2009, {\it \PR}D {\bf 80} 042005-1--7
	\bibitem{OcalPaper} Landry~M for the LIGO Scientific Collaboration 2005 {\it \CQG} {\bf 22} S985--94
	\bibitem{VirgoCal} Rolland~L 2009 to be submitted to {\it Amaldi 8 Conf. Proc.}
	\bibitem{GEOCal} Hewitson~M for the LIGO Scientific Collaboration 2007 {\it \CQG} {\bf 24} S445--55
	\bibitem{TamaCal1} Tatsumi~D and Tsunesada~Y and the TAMA Collaboration 2004 {\it \CQG} {\bf 21} S451--6
	\bibitem{LIGOhoft} Siemens~X, Allen~B, Creighton~J, Hewitson~M and Landry~M 2004 {\it \CQG} {\bf 21} S1723--36
	\bibitem{PcalTiming} Aso~Y, \etal 2009 {\it \CQG} {\bf 26} 055010 (13pp)
	\bibitem{Virgohoft} Marion~F, Mours~B and Rolland~L 2008 {\it Virgo doc.} VIR-078A-08 (https://tds.ego-gw.it)
	\bibitem{GEOhoft} Hewitson~M, Grote~H, Hild~S, L\"uck~H, Ajith~P, Smith~J~R, Strain~K~A, Willke~B and Woan~G 2005 {\it \CQG} {\bf 22} 4253--61
	\bibitem{TamaCal2} Telada~S, Tatsumi~D, Akutsu~T, Ando~M, Kanda~N and the TAMA Collaboration 2005 {\it \CQG} {\bf 22} S975--84
	\bibitem{S1paper} Adhikari R, Gonzalez G, Landry M and O'Reilly B (for the LIGO Scientific Collaboration) 2003 {\it \CQG} {\bf 20} S903--14
  \bibitem{LIGOPcal} Goetz~E, \etal 2009 accepted for publication in {\it \CQG} ({\it preprint} gr-qc/0910.5591)
  \bibitem{VCOpaper} Goetz~E and Savage~Jr~R~L 2009, in preparation for submission to {\it \CQG}
  \bibitem{PDHpaper} Drever~R~W~P, Hall~J~L, Kowalski~F~V, Hough~J, Ford~G~M, Munley~A~J and Ward~H 1983 {\it Appl. Phys.} B: {\it Lasers Opt.} {\bf 31} 97--105
  \bibitem{LIGOPDH} Regehr~M~W, Raab~F~J and Whitcomb~S~E 1995 {\it Opt. Lett.} {\bf 20} 1507--9
	\bibitem{GlasgowPcal} Clubley~D~A, Newton~G~P, Skeldon~K and Hough~J 2001 {\it \PL}A {\bf 283} 85--8
  \bibitem{GEOPcal} Mossavi~K, \etal 2005 {\it \PL}A {\bf 353} 1--3
  \bibitem{VirgoPcal} Rolland~L, Marion~F and Mours~B 2008 {\it Virgo doc.} VIR-053A-08 (https://pub3.ego-gw.it)
  \bibitem{GEOcompare} Hild S 2007, PhD Thesis University of Hannover
  \bibitem{HildEffect} Hild~S, \etal 2007 {\it \CQG} {\bf 24} 5681--8
  \bibitem{TMdeform} Badan~M~A, Landry~M, Savage~R and Willems~P 2009 {\it LIGO doc.} T0900401 (https://dcc.ligo.org)
  \bibitem{DynamResonance} Rakhmanov~M, Savage~Jr~R~L, Reitze~D~H and Tanner~D~B 2002 {\it \PL} A {\bf 305} 239--44
	\bibitem{S5paper} Kissel~J, \etal in preparation
\end{thebibliography}
\end{document}